\documentstyle[preprint,aps]{revtex}

\def\journal#1, #2, #3{
    {\sl #1~}{\bf #2} #3}
\def\pr{\journal Phys. Rev. }

\def\prl{\journal Phys. Rev. Lett. }

\def\np{\journal Nucl. Phys. }
\def\pl{\journal Phys. Lett. }

\def\mpl{\journal Mod. Phys. Lett. }

\def\jmp{\journal J. Math. Phys. }
\def\jp{\journal J. Phys. }
\def\beq{\begin{equation}}
\def\eeq{\end{equation}}
\def\ba{\begin{eqnarray}}
\def\ea{\end{eqnarray}}
\def\l{\lambda}
\def\o{\omega}
\def\f{\varphi}
\def\oo{\tilde{\omega}}
\def\et{\tilde{\eta}}
\def\r{\tilde{\rho_0}}
\def\G{\tilde{G}}
\def\k{\alpha}
\def\kx{\alpha(x-y)}
\newcommand{\cl}{Calogero}
\newcommand{\h}{Hamiltonian}
\newcommand{\col}{collective}
\newcommand{\pv}[1]{{-  \hspace {-4.0mm} #1}}

\begin{document}
\draft
\title{Collective-Field Excitations in the \cl\ Model}
\author{I. Andri\'c, V. Bardek and L. Jonke\footnote{e-mail address:
andric@thphys.irb.hr \\ \hspace*{3cm} bardek@thphys.irb.hr \\ \hspace*{3cm}
larisa@thphys.irb.hr }}
\address{Department of Theoretical Physics,\\
Rudjer Bo\v skovi\'c Institute, P.O. Box 1016,\\
41001 Zagreb, CROATIA}
\date{September, 1994}
\maketitle
\begin{abstract}
\widetext
We consider the large-N \cl\ model in the \h\ collective-field approach based
on the $1/N$ expansion. The Bogomol'nyi limit appears and the corresponding
equation for the semiclassical configuration gives the correct ground-state
energy. Using the method of the orthogonal polynomial we find the excitation
spectrum of density fluctuations around the semiclassical solution for any
value of the statistical parametar $\l$. The wave functions of the excited
states are explicitly constructed as a product of Hermite polynomials in terms
of the collective modes.
The two-point correlation function is calculated as a series expansion in
$1/\rho$ for any intermediate statistics.
\narrowtext
\end{abstract}
\vspace{1cm}
\pacs{03.65.Ss 05.50.-d 71.10.+x}
\newpage

Recently much attention has been paid to the quantum N-body \cl \ model in
$1+1$ space-time\cite{cal}. As is well known, the \cl\ model belongs to a large
family of one-dimensional quantum integrable models. In addition, it exhibits
fractional statistics, namely, its ground-state wave function is of a Jastrow
type and can be seen as the Laughlin wave function\cite{p1,sim}.
More recently, applying the collective-field approach to the fractional quantum
Hall droplet dynamics, it has been shown that the density correlation function
of the edge state interpolates to the correlation function of the \cl\
model\cite{iso}.
 Furthermore, the \cl\ model is closely related to the matrix models and
exhibits an exact equivalence with the collective-field formulation of the
$d=1$ string theory at the classical level\cite{j2}.
In addition, the collective-field theory with the cubic \h\ and other
higher-order terms in the $1/N$ expansion was recently used in deriving a new
interesting soliton solution in $1+1$ dimension\cite{j1}. This effective theory
can again be recognized as the $\l=1/2$ collective-field formulation of the
\cl\ model. It is, therefore, of considerable interest to study the general
features of the \cl\ model in a collective-field theoretical framework. To the
lowest order, the collective-field theory of the \cl\ model was shown to
correctly reproduce the first relevant term in the ground-state
energy\cite{a1,a2}.

In the present paper we extend these investigations by including higher-order
terms in the $1/N$ expansion. We also study the properties of the \cl\ model
concentrating on the excitation spectrum of small fluctuations around the
semiclassical configuration $\rho_0(x)$, the corresponding wave functions, and
the first quantum corrections (the one-loop contribution) to the ground-state
energy.

Before the studying of the excitation spectrum, let us briefly recall the
essential features of the collective-field approach to the Calogero
model\cite{a2}. The approach is described by a Hamiltonian
\ba \label{ha1}
H && =\frac{1}{2}\int dx\rho (x)(\partial _x\pi )^2+\frac{1}{2}\int dx\rho
(x)\left(\frac{\lambda -1}{2}\frac{\partial
_x\rho(x)}{\rho(x)}+\lambda\pv{\int} dy\frac{\rho(y)}{x-y}\right) ^2+
\nonumber \\
&& +\frac{\omega^2}{8}\int dxdy\rho(x)(x-y)^2\rho(y)-\frac{\lambda -1}{4}\int
dx\partial_x^2\delta (x-y)|_{y=x}-\frac{\lambda}{2}\int dx
\rho(x)\partial_x\frac{P}{x-y}|_{y=x}
,
\ea
where a dimensionless constant $\lambda$ determines the strength of the \cl\
pair coupling through the relation $\lambda (\lambda -1)=g$. $\omega $ is the
strength of a harmonic confinement potential.
The collective field $\rho(x)$ is the continuum limit of the dynamical quantity
\beq \label{r1}
\rho(x)=\sum_{i=1}^N\delta(x-x_i),
\eeq
where $x_i$ are the positions of N spinless bosonic particles. $\pi(x)$ is
canonical conjugate of the field $\rho(x)$:
\beq
[\partial_x\pi(x),\rho(y)]=-i\partial_x\delta(x-y)
.\label{r2}
\eeq
It follows from the definition (\ref{r1}) that the collective field obeys the
normalization condition
\beq \label{r3}
\int dx\rho(x)=N. \eeq
The first term in the collective \h , quadratic in the conjugate momentum
$\pi$, represents the kinetic energy of the system. The second term, rewritten
as a complete square, emerges as a quantum collective-field potential. The last
two terms represent a singular contribution, but, as will be shown later, they
are canceled by infinity zero-point fluctuations of the \col \ field $\rho$. To
find the ground-state energy and the corresponding \col \ motion of the \cl \
system in the large-N limit, we should minimize the energy functional
(\ref{ha1}) with respect to $\pi$ and $\rho$, obeying the normalization
condition (\ref{r3}). However, in our case, owing to the special features of
the model, there is a much more efficient method for doing the same thing.

Performing the $1/N$ expansion of the \col \ field $\rho(x)$ in the form\beq
\label{r5}
\rho(x)=\rho_0(x)+\eta(x), \eeq where $\rho_0(x)$ is the ground-state
semiclassical configuration and $\eta(x)$ a small density fluctuation around
$\rho_0(x)$
, we can rewrite the \col\  \h \ (\ref{ha1}) up to the quadratic terms in $\pi$
and $\eta$ as
\ba \label{ha2}
&& H=\frac{1}{2}\int dx\rho_0 (x)\left(\frac{\lambda -1}{2}\frac{\partial
_x\rho_0(x)}{\rho_0(x)}+\lambda\pv{\int}
dy\frac{\rho_0(y)}{x-y}-\omega\sqrt{\frac{N}{2}}x\right) ^2
+N\omega\sqrt{\frac{N}{2}}\left(\frac{\lambda N}{2}-\frac{\l -1}{2}\right)+
\nonumber \\
&& +\frac{1}{2}\int dx\rho_0(x)(\partial_x\pi)^2+\frac{\pi^2\l^2}{2}\int
dx\rho_0(x)\eta^2(x)+\frac{(\l -1)^2}{8}\int
dx\frac{(\partial_x\rho_0(x))^2}{\rho_0^3(x)}\eta^2(x)+  \nonumber \\
&& +\frac{(\l -1)^2}{8}\int
dx\partial_x\left(\frac{\partial_x\rho_0(x)}{\rho_0^2(x)}\right)\eta^2(x)
 +\frac{(\l -1)^2}{8}\int dx\frac{(\partial_x\eta(x))^2}{\rho_0(x)}+\frac{\l(\l
-1)}{2}\pv{\int} dxdy\frac{\partial_x\eta(x)\eta(y)}{x-y}+ \nonumber \\
&& +\frac{\o^2}{8}\int dxdy\eta(x)(x-y)^2\eta(y)-\frac{\lambda -1}{4}\int
dx\partial_x^2\delta (x-y)|_{y=x}-\frac{\lambda}{2}\int dx
\rho_0(x)\partial_x\frac{P}{x-y}|_{y=x}.
\ea
There are no terms in the \h \ (\ref{ha2}) linear in $\eta(x)$ as we expand
around the minimum of the dominant, large-N \col \ potential.

Owing to the positive definiteness of the first leading term, the Bogomol'nyi
limit appears. The  Bogomol'nyi bound is saturated by the positive normalizable
solution $\rho_0(x)$ of the equation
\beq
\frac{\l-1}{2}\frac{\partial_x\rho_0(x)}{\rho_0(x)}+\l\pv{\int}
dy\frac{\rho_0(y)}{x-y}=\o\sqrt{\frac{N}{2}}x, \label{r6} \eeq
with the ground-state energy equal to
\beq \label{e0}
E_0=\frac{\o}{2}\sqrt{\frac{N}{2}}[\l N(N-1)+N]. \eeq
We have not been able to obtain an analytic solution to this equation for
general $\l$. However, we can offer the asymptotic behavior of the
semiclassical configuration $\rho_0(x)$. For $\l\neq 1$, the expression for
$\rho_0(x)$ near the origin is
\beq \label{r7}
\rho_0(x)=A\exp \left[\frac{x^2}{\l-1}(\o\sqrt{\frac{N}{2}}+\l B)\right] , \eeq
where A and B are positive arbitrary constants. For $x\rightarrow\infty $, we
obtain
\beq \label{r8}
\rho_0(x)=Cx^{\frac{2\l
N}{1-\l}}\exp\left(\o\sqrt{\frac{N}{2}}\frac{x^2}{\l-1}\right) , \eeq
where C is again a positive constant. It is evident that the character of the
solution depends crucially on the value of the parametar $\l$. For $\l >1$, the
 solution $\rho_0$ exists on the compact support only.
As is well known, the coupling constant $\l$ specifies the statistics of the
\cl\ model. For special values of $\l$, i.e., $\l=0$ we have bosons and for
$\l=1/2,1,\;{\rm and}\;2$ the model is related to the system of orthogonal,
unitary (fermions) and symplectic matrix theory, respectively\cite{a2}. For the
bosonic and fermionic case, the equation (\ref{r6}) can be exactly solved. The
normalized bosonic distribution, $\l=0$, is given by \beq \label{r9}
\rho_0(x)=N\sqrt{\frac{\o}{\pi}\sqrt{\frac{N}{2}}}\exp\left(
-\o\sqrt{\frac{N}{2}}x^2\right) , \eeq
and has a tail, in contrast to the fermionic distribution, $\l=1$: \beq
\label{r10}
\rho_0(x)=\sqrt{\frac{\o}{\pi}\sqrt{\frac{N}{2}}\left(
2N-\o\sqrt{\frac{N}{2}}x^2\right)} , \eeq
which has a sharp boundary and is defined on the compact support only.

Let us now find an interesting \col-field hole excitation of the \cl \ model,
which can also be reached by the Bogomol'nyi saturation. Using the identity for
the principal distribution:
\beq \label{r18}
\frac{P}{x-y}\frac{P}{x-z}+\frac{P}{y-x}\frac{P}{y-z}+\frac{P}{z-x}\frac{P}{z-y}=\pi^2\delta(x-y)\delta(x-z), \eeq
and performing partial integration, we can rewrite the first, leading term in
the \h\ (\ref{ha2}) as
\ba \label{ha22}
&& \frac{1}{2}\int dx\rho_0 (x)\left(\frac{\lambda -1}{2}\frac{\partial _x\rho
_0(x)}{\rho_0(x)}+\lambda\pv{\int}
dy\frac{\rho_0(y)}{x-y}-\omega\sqrt{\frac{N}{2}}x\right) ^2=\nonumber \\
&& =\frac{1}{2}\int dx\rho_0 (x)\left(\frac{\lambda -1}{2}\frac{\partial _x\rho
_0(x)}{\rho_0(x)}+\lambda\pv{\int}
dy\frac{\rho_0(y)}{x-y}-\omega\sqrt{\frac{N}{2}}x+\frac{c}{x}\right)
^2-\nonumber \\
&& -\left(\frac{c^2}{2}+\frac{c(\l-1)}{2}\right)\int
dx\frac{\rho_0(x)}{x^2}+\omega cN\sqrt{\frac{N}{2}}+\frac{c\l}{2}\left(\int
dx\frac{\rho_0(x)}{x}\right)^2-\frac{c\l\pi^2}{2}\rho_0^2(0). \ea
For the symmetric configuration, $\rho_0(x)=\rho_0(-x)$ representing a hole
located at the origin, $\rho_0(0)=0$, and the particular value of the constant
$c$
given by $$ c=1-\l,$$ the Bogomol'nyi limit appears.
The contribution of the squared term in (\ref{ha22}) vanishes and the
corresponding configuration satisfies the enlarged Bogomol'nyi equation
\beq \label{b22}
\frac{\lambda -1}{2}\frac{\partial _x\rho_0(x)}{\rho_0(x)}+\lambda\pv{\int}
 dy\frac{\rho_0(y)}{x-y}-\omega\sqrt{\frac{N}{2}}x+\frac{1-\l}{x}=0 .\eeq
The role of the new, singular term in equation (\ref{b22}) is to compensate for
the
singularity produced by $\partial_x\ln \rho_0(x)$ at the origin, $x=0$.
The corresponding energy is given by
\beq \label{e22}
E=E_0+N\sqrt{\frac{N}{2}}\omega(1-\l), \eeq and can be lower than $E_0$ for
$\l>1$. A more detailed discusion of the Bogomol'nyi equation (\ref{b22}) will
be given elsewhere. From now on we restrict our investigation to the case of
the ground-state semiclassical configuration $\rho_0$ given by equation
(\ref{r6}).

To find the spectrum of low-lying excitations, we have to diagonalise part of
the \col\ \h \ (\ref{ha2}) quadratic in the operators $\pi$ and $\eta$. Let us
first eliminate $\rho_0(x)$ in the kinetic-energy part by introducing a
modified fluctuation $\et(x)$ through \beq \label{r11}
\eta(x)=\partial_x\left(\sqrt{\rho_0(x)}\et(x)\right) . \eeq
It is easy to see that in this case the kinetic energy transforms as
\beq \label{r12}
\frac{1}{2}\int dx\rho (x)(\partial _x\pi )^2\;\;\longrightarrow\;\;
-\frac{1}{2}\int dx\frac{\delta^2}{\delta\et^2(x)}. \eeq
Next we introduce the normal mode expansion of the fluctuation $\et$
\beq \label{r13}
\et(x)=\frac{1}{\sqrt{\rho_0(x)}}\sum_{n} q_n\f_n(x), \eeq
and the corresponding conjugate momentum \beq \label{r14}
\tilde{\pi}(x)=-i\frac{\delta}{\delta\et(x)}=\frac{1}{\sqrt{\rho_0(x)}}\sum_{n}
p_n\f_n(x), \eeq
where the operators $q_n$ and $p_n$ satisfy standard bosonic commutator algebra
\beq \label{r15}
[q_n,p_m]=i\delta_{nm},\;\;\; [q_n,q_m]=[p_n,p_m]=0. \eeq
Here, the yet unspecified functions $\f_n(x)$ form an orthonormal and complete
set in the sense that the following relations are satisfied:
\begin{mathletters}
\beq \label{n1}
\int dx\frac{\f_n(x)\f_m(x)}{\rho_0(x)}=\delta_{nm},
\eeq
\beq \label{n2}
\sum_{n} \frac{\f_n(x)\f_n(y)}{\sqrt{\rho_0(x)\rho_0(y)}}=\delta(x-y).
\eeq \end{mathletters}
Substituting the relations (\ref{r13}) and (\ref{r14}) into the \h\ (\ref{ha2})
and using (\ref{n1}), we obtain the \h\ in the diagonal form:
\begin{eqnarray}
 H && =E_0+\frac{1}{2}\sum_{n} p_n^2+\frac{1}{2}\sum_{n} \o_n^2q_n^2- \nonumber
\\
&& -\frac{\lambda -1}{4}\int dx\partial_x^2\delta
(x-y)|_{y=x}-\frac{\lambda}{2}\int dx \rho_0(x)\partial_x\frac{P}{x-y}|_{y=x},
\label{ha3} \end{eqnarray}
if the function $\f_n(x)$ satisfies the integro-differential equation
\ba \label{r16}
&&
\frac{(\l-1)^2}{8}\rho_0(x)\partial_x^2\left(\frac{\partial_x^2\f_n(x)}{\rho_0(x)}\right) -\frac{\pi^2\l^2}{2}\rho_0(x)\partial_x(\rho_0(x)\partial_x\f_n(x))- \nonumber \\
&&
-\frac{(\l-1)^2}{8}\rho_0(x)\partial_x\left\{\left[\frac{(\partial_x\rho_0(x))^2}{\rho_0^3(x)}+\partial_x\left(\frac{\partial_x\rho_0(x)}{\rho_0^2(x)}\right)\right] \partial_x\f_n(x)\right\} +  \nonumber \\
&& +\frac{\l(\l-1)}{2}\rho_0(x)\partial_x^2\pv{\int}
dy\frac{\partial_y\f_n(y)}{x-y}-\frac{\o^2}{4}\rho_0(x)\int
dy\f_n(y)=\frac{\o_n^2}{2}\f_n(x)
. \ea
It remains to prove that the corresponding eigenfrequencies $\o_n$ are given by
nonnegative numbers.

We can avoid the manipulation with this cumbersome equation if we realize that
it can be rederived by the simpler eigenequation
\ba \label{r17}
\o_n\f_n(x) && =\rho_0(x)\l\pv{\int}
dy\frac{\partial_y\f_n(y)}{x-y}+\rho_0(x)\frac{\l-1}{2}\partial_x\left(\frac{\partial_x\f_n(x)}{\rho_0(x)}\right) - \nonumber \\
&& -\frac{\o}{\sqrt{2N}}\rho_0(x)\int dy\f_n(y). \ea
Substituting the  expression (\ref{r17}) for $\f_n$ into the right-hand side of
the same expression and using the identity for the principal distributions
(\ref{r18})
we easily obtain the eigenequation (\ref{r16}).
The solutions to the equation (\ref{r17}) indeed form an orthogonal set of
functions on the interval $-\infty <x<\infty $. To prove the orthogonality
(\ref{n1}), we write the equation for $\f_n(x)$, multiply by $\f_m(x)$, and
then integrate over the interval
\ba \label {n3}
&& \o_n\int\frac{\f_n(x)\f_m(x)}{\rho_0(x)}dx=\l\pv{\int}
dydx\frac{\partial_y\f_n(y)\f_m(x)}{x-y}+ \nonumber \\
&& +\frac{\l-1}{2}\int
dx\f_m(x)\partial_x\left(\frac{\partial_x\f_n(x)}{\rho_0(x)}\right)
-\frac{\o}{\sqrt{2N}}\int dx\f_n(x)\int dy\f_m(y). \ea
If we now write (\ref{n3}) with n and m interchanged, subtracting it from
(\ref{n3}), and integrating by parts, we obtain the orthogonality condition
\beq \label{n4}
(\o_n-\o_m)\int dx\frac{\f_n(x)\f_m(x)}{\rho_0(x)}=0. \eeq
For $n\neq m$, the integral must vanish. Technically, the proof of the
completeness of the set $\{\f_n(x)\}$ in (\ref{n2}) is outside the scope of the
present paper. We therefore omit it and simply anticipate that the solutions to
the equation (\ref{r17}) really satisfy the closure relation (\ref{n2}).

We are now in a position to calculate the inverse of the density-density
correlation function in the \cl\ model. The vacuum state of the operator part
of the \h\ (\ref{ha3}) is given by \beq \label{r19}
|0\rangle =\exp\left( -\frac{1}{2}\sum_{n}\o_nq_n^2\right) .\eeq
This may be reexpressed in terms of $\et(x)$ using an expression inverse to
(\ref{r13}) \beq \label{r20}
q_n=\int dx\et(x)\frac{\f_n(x)}{\sqrt{\rho_0(x)}}, \eeq
 and further simplified using the  eigenequation (\ref{r17}). We finally obtain
\beq \label{r21}
|0\rangle =\exp\left( -\frac{1}{4}\int dxdy\eta(x)G^{-1}(x,y)\eta(y)\right),
\eeq
where the inverse of the correlation function is given by \beq \label{r22}
G^{-1}(x,y)=-2\l\ln
|x-y|-(\l-1)\frac{\delta(x-y)}{\rho_0(x)}-\o\sqrt{\frac{2}{N}}xy. \eeq

By functional integration over $\eta$ we can easily check that $G^{-1}(x,y)$
indeed represents the inverse of the correlation function $G(x,y)$:
\ba \label{r23}
&& \langle 0|\rho(x)\rho(y)|0\rangle =\rho_0(x)\rho_0(y)+\langle
0|\eta(x)\eta(y)|0\rangle = \nonumber \\ && =\rho_0(x)\rho_0(y)+\frac{\int{\cal
D}\eta\eta(x)\eta(y)\exp\left( -\frac{1}{2}\int \eta
G^{-1}\eta\right)}{\int{\cal D}\eta\exp\left( -\frac{1}{2}\int \eta
G^{-1}\eta\right)} = \nonumber \\ && =\rho_0(x)\rho_0(y)+G(x,y), \ea
where we have used the fact that the vacuum expectation value of the
fluctuation $\eta$  vanishes.

Before proceeding we should point out that $G^{-1}(x,y)$ is determined up to
the symmetric combination $f(x)+f(y)$ where $f(x)$ is an arbitrary real
function. This is because of the normalization condition (\ref{r3}), the
expansion (\ref{r5}), and the symmetry structure of $G$ and $G^{-1}$
(\ref{r22}).

According to the expression for the correlation function $G(x,y)$
 (\ref{r23}), we can rewrite the eigenequation (\ref{r17}) as
\beq \label{y1}
\o_n\f_n(x)=\frac{1}{2}\rho_0(x)\int dy\f_n(y)\partial_x\partial_y G^{-1}(x,y)
.\eeq
Multiplying it by $\f_n(z)$ and summing over $n$ we obtain
\beq \label{y2}
G(x,y)=\frac{1}{2}\partial_x\partial_y\sum_n\frac{\f_n(x)\f_n(y)}{\o_n}. \eeq
Equations (\ref{y2}) and (\ref{r17}) imply that the correlation function
$G(x,y)$ satisfies the following integro-differential equation:
\beq \label{xx1}
2\l\pv{\int}\frac{dyG(y,z)}{x-y}+(\l-1)\partial_x\left(\frac{G(x,z)}{\rho_0(x)}\right) =\partial_z\delta (x-z) .\eeq
We can covert this equation into the differential equation using the identity
for the principal distribution (\ref{r18}),
\beq \label{xx2}
G(x,y)=-\frac{\l-1}{4\l^2\pi^2}\partial_x\left[\frac{1}{\rho_0(x)}\partial_x
\left((\l-1)\frac{G(x,y)}{\rho_0(x)}+\delta(x-y)\right)\right]+
\frac{1}{2\l\pi^2\rho_0(x)}\partial_y\left(\frac{\rho_0(y)}{y-x}\right) . \eeq
By rescaling the \col \ field and the correlation function as
\begin{mathletters}
\beq \rho_0(x)=\frac{\l-1}{\l}\r(x), \eeq
\beq G(x,y)=\frac{1}{\l}\G(x,y) , \eeq \end{mathletters}
and iterating equation (\ref{xx2}) (with respect to $G(x,y)$), we can compute
the correlation function in terms of $\rho_0(x)$ to arbitrary order:
\ba \label{xx3}
\G(x,y) && =\frac{1}{2\pi^2\r(x)}\partial_y\left(\frac{\r(y)}{y-x}\right)-
\frac{1}{4\pi^2}\partial_x\left(\frac{1}{\r(x)}\partial_x\delta(x-y)\right)
\nonumber \\ &&
-\frac{1}{8\pi^4}\partial_x\left\{\frac{1}{\r(x)}\partial_x\left[
\frac{1}{\r^2(x)}\partial_y\left(\frac{\r(y)}{y-x}\right)\right]\right\}
\nonumber \\ &&
+\frac{1}{16\pi^4}\partial_x\left\{\frac{1}{\r(x)}\partial_x\left[
\frac{1}{\r(x)}\partial_x\left(\frac{1}{\r(x)}\partial_x\delta(x-y)\right)\right]\right\} + \cdots . \ea
Equation (\ref{xx3}) is basically the $1/\rho_0$ expansion and we anticipate
that it converges. We stress that expresion (\ref{xx3}) holds for $\l$
different
from zero and one, i.e. for generic intermediate statistics.

At this point we would like to analyse the structure of the static correlation
function $G(x,y)$ in the reduced \cl \ model, i.e. in the model without
confining harmonic interaction $(\o=0)$. It is easy to see that, in this case,
the solution to equation (\ref{r6}) is given by the constant-density
configuration
$\rho=\rho_0$. It can be readily shown that $G(x,y)$ satisfies the following
second-order differential equation
\beq \label{xx4}
\left(\partial_x^2+\frac{4\l^2\pi^2\rho_0^2}{(\l-1)^2}\right) G(x,y)=
-\frac{\rho_0}{\l-1}\partial_x^2\delta(x-y)+\frac{2\l\rho_0^2}{(\l-1)^2}
\partial_x\left(\frac{1}{x-y}\right) . \eeq
Having in mind the translation-invariance of the correlation function, let us
now rewrite equation (\ref{xx4}) in momentum space by Fourier transforming
 the function $G(x-y)$
\beq \label{xx5}
G(x-y)=\frac{\rho_0}{2\pi}\int dk e^{ik(x-y)}\G(k) , \eeq
and the principal-value distibution
\beq \label{xx6}
\frac{P}{x-y}=\frac{1}{2i}\int dke^{ik(x-y)}{\rm sign}k , \eeq
\beq \label{xx7}
\left(\frac{4\l^2\pi^2\rho_0^2}{(\l-1)^2}-k^2\right) \G(k)=\frac{k^2}{\l-1}+
\frac{2\pi\l\rho_0}{(\l-1)^2}|k|. \eeq
We note that for $0<\l<1$, the correlation in the momentum space $\G(k)$ is
positive and depends only on the absolute value of $k$. It can be written in
the
form \beq \label{xx8} \G(k)=\frac{k^2}{2\o(k)} , \eeq
with the dispersion $\o(k)$ given by
\beq \label{xx80}
\o(k)=\l\pi\rho_0|k|-\frac{\l-1}{2}k^2 . \eeq
Then we obtain the expression for the correlation function $G(x-y)$
\beq \label{xx9}
G(x-y)=\frac{\rho_0}{2\pi(\l-1)}\int dk\frac{e^{ik(x-y)}|k|}{\frac{2\l}{\l-1}
k_f-|k|} , \eeq with $k_f=\pi\rho_0$, representing the Fermi momentum. The
 integral can be recast as follows:
\beq \label{xx10}
G(x-y)=-\frac{\rho_0}{\l-1}\delta(x-y)+\frac{2\l\rho_0^2}{(\l-1)^2}\int_0^{\infty} dk\frac{\cos k(x-y)}{\frac{2\l}{\l-1}k_f-|k|}. \eeq
Using the Table of integrals\cite{g}, the correlation $G(x-y)$ turns out to be
\ba \label{xx11}
&& G(x-y)=-\frac{\rho_0}{\l-1}\delta(x-y)+\nonumber \\
        &&+\frac{2\l\rho_0^2}{(\l-1)^2}\left[
ci(\kx)\cos(\kx)+si(\kx)\sin(\kx)\right] , \ea
where $\k=2\l k_f/(\l-1)$ and $ci(x)\; (si(x))$ denotes the sine (cosine)
integral functions, respectively. In the fermionic case $\l=1$, the correlation
function $G(x,y)$ given by relation (\ref{xx2}) reduces to
\beq
G(x-y)=-\frac{1}{2\pi^2}\frac{1}{(x-y)^2} , \eeq
which in momentum space agrees with the correlation function found by the
authors of the reference\cite{ma} in the $|k|<2k_f$ sector. In our approach,
the other sector $|k|>2k_f$ is absent because of the large-N limit
$(k_f=\pi\rho_0\longrightarrow\infty)$. In the $\l=1/2(2) $ case, by expanding
the correlation function (\ref{xx8}) in the powers of $|k|/k_f$, up to the
cubic terms, we easily obtain the result of the reference\cite{ma}, again in
the $|k|<2k_f(4k_f)$ sector only.

Turning back to the \cl \ model with confining interaction, let us show that
there is a cancelation between the divergent vacuum energy $1/2\sum_n\o_n$ of
the harmonic \h\ in (\ref{ha3}) and the divergent last two terms. To this end,
we compute the vacuum energy as \beq \label{e1}
E_{vac}=\frac{\langle 0|H^{(2)}|0\rangle}{\langle 0|0\rangle}, \eeq
but this time with the vacuum $|0\rangle$ and the harmonic part $H^{(2)}$ of
the \h\ (\ref{ha3}) given in terms of the density fluctuation $\eta(x)$. Using
equation (\ref{r6}) for $\rho_0(x)$, we finally obtain \beq \label{e2}
E_{vac}=\frac{\lambda -1}{4}\int dx\partial_x^2\delta
(x-y)|_{y=x}+\frac{\lambda}{2}\int dx \rho_0(x)\partial_x\frac{P}{x-y}|_{y=x}
-\frac{\o}{2}\sqrt{\frac{N}{2}}. \eeq
Hence the total \h\ becomes \beq \label{e3}
H=\frac{\o}{2}\sqrt{\frac{N}{2}}[\l N(N-1)+N-1]+\sum_{n}\o_na_n^{\dag}a_n, \eeq
where we have introduced the standard creation and annihilation bosonic
operators
\begin{mathletters}
\beq \label{a}

a_n^{\dag}=\frac{1}{\sqrt{2}}\left(\sqrt{\o_n}q_n-\frac{i}{\sqrt{\o_n}}p_n\right), \eeq
\beq \label{a1}
a_n=\frac{1}{\sqrt{2}}\left(\sqrt{\o_n}q_n+\frac{i}{\sqrt{\o_n}}p_n\right).
\eeq \end{mathletters}
The ground-state energy obtained in (\ref{e3}) coincides with the \cl\ results
\cite{cal}. To find the eigenfrequencies $\o_n$, let us for the moment suppose
that the solution $\f_n(x)$ of the eigenequation (\ref{r17}) can be cast into
the form \beq \label{r24}
\f_n(x)\sim\rho_0(x)x^n , \eeq
where n is a nonnegative integer. In this case we can evaluate the integral on
the right-hand side of the eigenequation (\ref{r17}) using the formula \beq
\label{r25}
\frac{x^n-y^n}{x-y}=x^{n-1}+x^{n-2}y+\cdots +xy^{n-2}+y^{n-1}. \eeq
Introducing the moments $m_k$ of the distribution function $\rho_0(x)$
\beq \label{r26}
m_k=\int dxx^k\rho_0(x) , \eeq
and using the equation (\ref{r6}) for $\rho_0(x)$, we can reduce the right-hand
side
of the equation (\ref{r17}) to \ba \label{r27}
&&
\o\sqrt{\frac{N}{2}}(n+1)\rho_0(x)x^n+\rho_0(x)x^{n-2}\left[\frac{\l-1}{2}n(n-1)-\l(n-1)m_0\right] - \nonumber \\ && -\l\rho_0(x)x^{n-4}(n-3)m_2+
\cdots + \l\rho_0(x)m_{n-2}-\frac{\o}{\sqrt{2N}}m_n\rho_0(x).
\ea
The structure of this expression indicates that the exact eigenfunction
$\f_n(x)$ is given by an appropriate linear combination of different power of
$x$:
\beq \f_n(x)=\rho_0(x)\sum_{p=0}^{n}c_p^nx^p. \eeq
Substituting this expression into the equation (\ref{r17}) and matching the
coefficients of each power of $x$, we obtain an algebraic system of $n+1$
homogeneous equations in $c_p^n$:
\ba \label{det}
&& c_n^n(n+1-\oo_n)=0,  \nonumber \\
&& c_{n-1}^n(n-\oo_n)=0,  \nonumber \\
&& c_{n-2}^n(n-1-\oo_n)+c_n^n\left(\frac{\l-1}{2}n(n-1)-\l m_0(n-1)\right)=0,
\nonumber \\
&& c_{n-3}^n(n-2-\oo_n)+c_{n-1}^n\left(\frac{\l-1}{2}(n-1)(n-2)-\l
m_0(n-2)\right)=0,  \nonumber \\
&& c_{n-4}^n(n-3-\oo_n)+c_{n-2}^n\left(\frac{\l-1}{2}(n-2)(n-3)-\l
m_0(n-3)\right)-\l c_n^nm_2(n-3)=0,  \nonumber \\
&& \vdots \nonumber \\
&& c_2^n(3-\oo_n)+6(\l-1)c_4^n-3\l\sum_{k=0}^{n'}c_{2k+4}^nm_{2k}=0,  \nonumber
\\
&& c_1^n(2-\oo_n)+3(\l-1)c_3^n-2\l\sum_{k=0}^{n'}c_{2k+3}^nm_{2k}=0,  \nonumber
\\
&&
c_0^n(1-\oo_n)+(\l-1)c_2^n-\l\sum_{k=0}^{n'}c_{2k+2}^nm_{2k}-\frac{1}{N}\sum_{k=0}^{n'+1}c_{2k}^nm_{2k}=0,
\ea
where $\o_n=\oo_n\;\o\sqrt{N/2}$, and $n'$ is
\[n'=\left\{ \begin{array}{l}
   \frac{n-2}{2}\;\;\;\;{\rm for\;\; n\;\;even } \\
   \frac{n-3}{2}\;\;\;\; {\rm for\;\;n\;\;odd}.  \end{array}
\right. \]
The moments $m_k$, for k odd, vanish because the distribution $\rho_0(x)$ is a
symmetric function.
A system of homogeneous equations has a nontrivial solution if its determinant
vanishes.
 The determinant of our system (\ref{det}) is just a product of diagonal terms
(matrix of our system of equations is an upper-triangle matrix)
\beq \label{d}
D=(\oo_n-n-1)(\oo_n-n)(\oo_n-n+1)\cdots(\oo_n-3)(\oo_n-2)\oo_n , \eeq so it is
easy to see that the eigenfrequencies $\o_n$ are
\beq \label{o}
\o_n =\left\{ \begin{array}{l}
0\;\;\;{\rm for\;\;n=0} \\
(n+1)\;\o\sqrt{\frac{N}{2}}\;\;\;{\rm for\;\;n\neq0}. \end{array} \right. \eeq

The excitation spectrum of the \cl \ model is thus determined by the
diagonalised \h\
\beq \label {ha5}
H_{exc}=\sum_{n=1}^{\infty}\o\sqrt{\frac{N}{2}}(n+1)a_n^{\dag}a_n. \eeq
The general excited-state vector can be built by repeatedly acting with the
creation operator (\ref {a}) on the vacuum state:
\beq \label {ex1}
S\left \{\prod_n^{\infty}(a_n^{\dag})^{g_n}\right\}|0\rangle\sim S\left
\{\prod_n^{\infty}H_{g_n}(\sqrt{\o_n}q_n)\right\}|0\rangle .  \eeq
The integer $g_n$ represents the occupation numbers of the nth oscillator mode,
$H_{g_n}$ is the Hermite polynomial of order $g_n$ and $S$ denotes total
symmetrization. The corresponding excitation energy is \beq \label{ex2}
E_{\{g_n\}}=\o\sqrt{\frac{N}{2}}\sum_{n=1}^{\infty}(n+1)g_n.
\eeq
This is again in complete accordance with the \cl \ result, the only difference
being in $N\rightarrow\infty$.
It is interesting to note that the excitation spectrum (\ref{ex2}) does not
depend on the statistical parametar $\l$.

Let us now list the first few eigenfuctions $\f_n(x)$ and the corresponding
eigenfrequencies:
\ba \label{ex3}
&& \f_0(x)=c_0^0\rho_0(x),\;\;\;\o_0=0, \nonumber\\
&& \f_1(x)=c_1^1x\rho_0(x),\;\;\;\o_1=2\o\sqrt{\frac{N}{2}},\nonumber\\
&&
\f_2(x)=(c_0^2+c_2^2x^2)\rho_0(x),\;\;\;\o_2=3\o\sqrt{\frac{N}{2}},\nonumber\\
&&
\f_3(x)=(c_1^3x+c_3^3x^3)\rho_0(x),\;\;\;\o_3=4\o\sqrt{\frac{N}{2}},\nonumber\\
&&
\f_4(x)=(c_0^4+c_2^4x^2+c_4^4x^4)\rho_0(x),\;\;\;\o_4=5\o\sqrt{\frac{N}{2}},\nonumber \\
&&
\f_5(x)=(c_1^5x+c_3^5x^3+c_5^5x^5)\rho_0(x),\;\;\;\o_5=6\o\sqrt{\frac{N}{2}},\nonumber\\
&& \vdots  \ea
The fact that equation (\ref{r17}) has an eigensolution $\f_0(x)$ with
vanishing eigenvalue $\o_0$ is a consequence of the translational invariance of
the \cl \ model.
We note that the polynomial structures in $\f_n(x)$ have only even powers of
$x$ or only odd powers of $x$, depending on $n$ being even or odd.
The coefficients $c_p^n$ are interrelated by the system (\ref{det}) and the
free ones can be determined from the normalization condition (\ref{n1}). For
example, \ba \label{ex4}
&& 3\o\sqrt{\frac{N}{2}}c_0^2= c_2^2(-\l
N+\l-1-\o\frac{m_2}{\sqrt{2N}}),\nonumber\\ &&
2\o\sqrt{\frac{N}{2}}c_1^3=c_3^3(-2\l N+3(\l-1)),\nonumber\\
&& 2\o\sqrt{\frac{N}{2}}c_2^4=c_4^4(-3\l N+6(\l-1)),\nonumber\\&&
5\o\sqrt{\frac{N}{2}}c_0^4=c_4^4(-\l m_2-\o\frac{m_4}{\sqrt{2N}})+c_2^4(-\l
N+\l-1-\o\frac{m_2}{\sqrt{2N}}),\nonumber \\
&& 2\o\sqrt{\frac{N}{2}}c_3^5=c_5^5(-4\l N+10(\l-1)) ,\nonumber\\
&& 4\o\sqrt{\frac{N}{2}}c_1^5=c_3^5(-2\l N+3(\l-1))-c_5^52\l m_2, \nonumber\\
&& \vdots  \ea
All moments $m_k$ can be evaluated using the recurrence relation
\beq \label{ex5}
m_k=\frac{1}{\o}\sqrt{\frac{2}{N}}\left\{\left[ -(\l-1)\frac{k-1}{2}\right]
m_{k-2}+\l\sum_{i=1}^{k/2-1}m_{k-2i}m_{2i-2} -\frac{\l}{2}m_{k/2-1}^2 \right\}
,\eeq
which follows from equation (\ref{r6}) and formula (\ref{r25}). Here we list
several few moments:
\ba \label{ex6}
&& m_0=N ,\nonumber \\
&& m_2=\frac{1}{\o}\sqrt{\frac{N}{2}}(\l N-\l+1), \nonumber \\
&& m_4=\frac{1}{\o^2}\left[\l N-\frac{3}{2}(\l-1)\right] (\l N-\l+1). \ea
We can go on with this procedure and it is quite obvious that a systematic
construction of all polynomial solutions to the eigenequation (\ref{r17}) can
be accomplished.

Having discussed the eigenfunctions and the corresponding eigenvalues of
equation (\ref{r17}) for the general value of the statistical parametar $\l$,
we should briefly analyze the boson ($\l$=0) and the fermion ($\l$=1) case
separately. For $\l$=0, using the explicit form for $\rho_0$ (\ref{r9}), it is
easily verified that the eigenequation (\ref{r17}) reduces to the differential
equation for the Hermite polynomials. So $\f_n$ is given by \beq \label{b}
\f_n(x)=\rho_0(x)H_n\left(\sqrt{\o\sqrt{\frac{N}{2}}}\;x\right) ,\;\;n\geq
1,\eeq
which can be easily recovered from the set (\ref{ex4}). This is in accordance
with the result obtained by Jevicki and Sakita in\cite{bos}, for a large number
of harmonic oscillators. For the fermionic case (or the matrix model), $\l=1$,
the equation effectively reduces to the well-known formula for the Hilbert
transform of the Chebyshev polynomials $U_n$ and $T_n$ \beq \label{ch}
\pv{\int}_{-1}^{+1}\;\frac{T_n(y)dy}{(y-x)\sqrt{1-y^2}}=\pi U_{n-1}(x) . \eeq
Using the explicit form for $\rho_0$ (\ref{r10}), one can show that $\f_n$
reduces to \beq \label{f}
\f_n(x)=\sin \left(\frac{n\pi}{T}t(x)\right)
=\rho_0(x)U_n\left(\sqrt{\frac{\o}{2}\sqrt{\frac{2}{N}}}\;x\right),\;\;n\geq 1,
\eeq
where $t(x)$ is the time of flight of a classical particle in the harmonic
potential, with $T$ being the semiperiod. This is in exact agreement with the
calculations given in\cite{fer}. It can again be verified that our set
(\ref{ex4}) correctly reproduces the coefficients for the Chebyshev polynomials
$U_n$.
Hence, our set of orthogonal polynomials correctly interpolates between the two
well-known cases of bosons and fermions and, for intermediate values of $\l$,
gives the excitation spectrum and the corresponding normal modes for particles
with fractional statistics.

It is interesting that the fermionic case $(\l=1)$ is closely related to the
random-matrix theory which does not possess kinetic term $\partial_x\rho /\rho$
in
the \col\ \h \ (\ref{ha1}). This theory describes dymanics of the eigenvalues
of the orthogonal, unitary and symplectic matrices.
Recently, the authors of reference\cite{bz} reported a simple result for the
correlation of the eigenvalue density at two points in the spectrum. Now, we
are going to show that our correlation function $G(x,y)$, calculated up to the
quadratic term in the fluctuations around the semiclassical configuration
$\rho_0(x)$, is in agreement with the Br\'ezin-Zee spectral correlator given in
\cite{bz}.
Having in mind the relations for $G(x,y)$ (\ref{y2}), $\f_n(z)$ (\ref{f}), and
$\o_n$ (\ref{o}),
we easily obtain
\beq \label{y3}
G(x,y)=\frac{T}{4\pi^2}\frac{1}{\rho_0(x)\rho_0(y)}\partial_t\partial_{t'}
\left(\sum_n\frac{\cos \frac{n\pi}{T}(t-t')}{n}-\sum_n\frac{\cos
\frac{n\pi}{T}(t+t')}{n}\right) .\eeq
Next we employ the summation formula\cite{g}
\beq
\sum_n\frac{\cos n\alpha}{n}=-\frac{1}{2}\ln 2(1-\cos\alpha), \eeq
so that $G(x,y)$ can be rewritten as
\beq \label{y4}
G(x,y)=-\frac{1}{4T\rho_0(x)\rho_0(y)}\frac{1-\cos \frac{\pi}{T}t(x)\cos
\frac{\pi}{T}t(y)}{\left(\cos \frac{\pi}{T}t(x)-\cos
\frac{\pi}{T}t(y)\right)^2}.\eeq
In our case (\ref{r10}) the end points of the spectrum are given by $-a$ and
$a$, and the time of flight than reads
\beq \label{y5}
t(x)=\int_{-a}^x\frac{dy}{\rho_0(y)}=\int_{-a}^x\frac{dy}{\frac{\pi}{T}\sqrt
{a^2-y^2}}=\frac{T}{\pi}\left( \arcsin\frac{x}{a}+\frac{\pi}{2}\right) . \eeq
Inverting this relation, we obtain
\beq \label{y6}
x=-a\cos\frac{\pi}{T}t(x). \eeq Combination of equations (\ref{y4}) and
(\ref{y6}) yields the exact two-point correlation function of
reference\cite{bz}
\beq \label{y7}
G(x,y)=-\frac{1}{4T\rho_0(x)\rho_0(y)}\frac{a^2-xy}{(x-y)^2}, \eeq
valid for any even potential in which the eigenvalues `move`.
We note in passing that this result can be rederived from equation (\ref{xx2})
simply by putting $\l=1$.

Finally, it would be interesting to expand the collective \h\ (\ref{ha1}) up to
the cubic terms in the fluctuations around the semiclassical configuration
$\rho_0(x)$. These cubic interaction terms would provide a basis for systematic
perturbative computations of scattering amplitudes, higher (loop) corrections
to the ground-state energy and the dispersion law. These and other related
issues are currently under consideration.
\acknowledgments

This work was supported by the Scientific Fund of the Republic of Croatia.

\end{document}